\documentclass[aps,preprintnumbers,nofootinbib,superscriptaddress,showpacs,letterpaper,twocolumn]{revtex4-1}
\pdfoutput=1
\usepackage{amsmath,amssymb,amsbsy}
\usepackage[utf8]{inputenc}
\usepackage{graphicx}
\usepackage{ulem,color}
\usepackage[colorlinks=true, linkcolor=blue, filecolor=blue, urlcolor=blue, citecolor=blue, pdftex, plainpages=false]{hyperref}

\newcommand\buaff{Department of Physics and Center for Computational Science, Boston University, Boston, MA 02215, USA}
\newcommand\coaff{Department of Physics, University of Colorado, Boulder, CO 80309, USA}
\newcommand\Higgsaff{Higgs Centre for Theoretical Physics, School of Physics \& Astronomy, The University of Edinburgh, Edinburgh, EH9 3FD, UK}

\newcommand{\eq}[1]{Eq.~(\ref{#1})}
\newcommand{\fig}[1]{Fig.~\ref{#1}}

\definecolor{cyan}{rgb}{0, 0.55, 0.55}
\definecolor{orange}{rgb}{1.0, 0.5, 0}
\definecolor{Green}{rgb}{0, 0.588, 0}

\begin{document}
\preprint{EDINBURGH 2016/15}

\title{Large scale separation and hadronic resonances from a new strongly interacting sector}
\author{A.~Hasenfratz}\affiliation\coaff
\author{C.~Rebbi}\affiliation\buaff
\author{O.~Witzel}\affiliation\Higgsaff
\date{\today}

\begin{abstract}

Many theories describing physics beyond the Standard Model
rely on a large separation of scales.  Large scale separation arises in
models with mass-split flavors if the system is conformal in the
ultraviolet but chirally broken in the infrared.  Because of the
conformal fixed point, these systems exhibit hyperscaling and a highly
constrained resonance spectrum. We derive hyperscaling relations and
investigate the realization of one such system with four light and eight
heavy flavors. Our numerical simulations confirm that
both light-light and heavy-heavy resonance masses show hyperscaling and depend
only on the ratio of the light and heavy flavor masses.  The heavy-heavy spectrum is
qualitatively different from QCD and exhibits quarkonia with masses not 
proportional to the constituent quark mass. These resonances are only a few
times heavier than the light-light ones, which would put them within reach of
the LHC.
\end{abstract}
\maketitle

\section{Introduction}

 Various analysis of data accumulated by ATLAS and CMS in 2016 were recently presented at ICHEP (see e.g.~\cite{ICHEP2016_LHC_highlights1, ICHEP2016_LHC_highlights2}), yet these analysis do not reveal strong evidence for beyond-Standard Model (BSM) phenomena. Nevertheless the Standard Model (SM) is undoubtedly only an effective model.
New interactions are necessary to avoid triviality of the scalar sector,  describe dark matter or explain neutrino physics, UV complete the Higgs sector, etc. Many  viable BSM models rely on large scale separation between the infrared (IR) and ultraviolet (UV) physics \cite{Contino:2010rs,Luty:2004ye,Dietrich:2006cm,Luty:2008vs,Brower:2015owo,Csaki:2015hcd,Arkani-Hamed:2016kpz}. 
Such a scenario naturally leads to a ``walking'' gauge coupling and provides a dynamical mechanism for electroweak (EW) symmetry breaking, avoiding  unnaturally large tuning of the Higgs mass, while satisfying  stringent EW precision measurement constraints.

A possibility to achieve large scale separation is to add fermions that are not or only partially gauged under the  SM. In this paper we investigate  non-perturbative properties of  one  specific realization. In the UV we start with a gauge system and $N_f$   flavors within the conformal window and drive the system into the chirally broken  regime by lifting the mass of some  flavors.  The energy scale of chiral symmetry breaking is determined by the mass of the heavy flavors.  We do not explore the mechanism generating the mass of the heavy flavors but note that by  tuning that mass  the UV and IR scales can be separated arbitrarily. The UV dynamics is dominated by the  conformal infrared fixed point (IRFP)  guaranteeing  hyperscaling and high  predictability of the resonance spectrum made up of both light and heavy flavors. 
In Reference \cite{Brower:2015owo} we investigated such a system with $N_f=12$ fermions, splitting the masses into $N_\ell=4$ light and $N_h=8$ heavy flavors. We showed that if the system in the  $m_\ell = m_h =0$ limit is conformal, it exhibits hyperscaling in $m_h$  in the $m_\ell =0$ chiral limit. Our numerical results verified this expectation.  We also found that the light resonance spectrum of the $4\ell +8h$ system contained a relatively light $0^{++}$ state (i.e.~significantly lighter than the vector meson), while the rest of the masses interpolated between the $N_f=12$ and $N_f=4$ limits. Further details on the lattice implementation and simulations can be found in~\cite{Brower:2015owo,Hasenfratz:2016uar,Longpaper_inprep}.

Here we significantly extend our understanding of the $N_f=4\ell +8h$ model.  In Sect.~\ref{sec:hyperscaling} we   generalize the original derivation of Ref.~\cite{Brower:2015owo} and   deduce   that  universal hyperscaling should hold  in the light-light, heavy-light,  and heavy-heavy  sector not only in the $m_\ell=0$ chiral limit but  more generally as  function of  $m_\ell/m_h$.  These results describe general properties of  quantum field theories that are defined in the basin of attraction of a conformal IRFP.  In particular, ratios  of hadron masses or hadron mass over decay constants as  function of $m_\ell/m_h$  are expected to follow a  common functional form, independent of the individual values of the masses.  In the basin of attraction of the conformal fixed point, the gauge coupling is irrelevant and the infinite cutoff continuum limit is reached as $m_h\to 0$. Physical predictions  are independent of the gauge coupling when  $m_h$ goes to zero.

In Sect.~\ref{sec:numres} we verify these expectations in numerical simulations and show that  dimensionless ratios depend on $m_\ell/m_h$  even when varying the gauge coupling. We find that the heavy-heavy resonance states  remain relatively light, only a couple of times heavier than the light-light spectrum.    This feature is very different from QCD where the heavy-heavy spectrum of the strange, charm, or bottom mesons depend strongly on the  quark masses. 
 Depending on how the SM is coupled to the  IR system, the resonance states of the $N_f=4+8$ model could be in the few TeV range and therefore accessible at the LHC.

There are many phenomenological models that can be described by systems similar to ours.  The first ``walking'' models emerged in the context of technicolor theories and assumed that the new BSM physics is described by a near-conformal but chirally broken  gauge-fermion system. The massless pions  couple to the SM fields and break electroweak symmetry, while the Higgs boson might emerge  as a regular $0^{++}$ bound state of the system  \cite{Weinberg:1979bn,Susskind:1978ms,Eichten:1979ah,Holdom:1984sk,Yamawaki:1985zg,Bando:1987br,Appelquist:1991nm}. It is far from certain that any system is close enough to the conformal window to exhibit the necessary ``walking'' behavior and large mass anomalous dimension, yet there is indication from lattice studies that near-conformal models can have light  $0^{++}$ states, often referred to as ``dilaton-like'' Higgs bosons, as they might emerge from a broken scale symmetry~\cite{Aoki:2014oha,Aoki:2013xza, Appelquist:2016viq,Fodor:2014pqa,Fodor:2016wal,DeGrand:2015zxa,Nogradi:2016qek}. 
Composite Higgs models offer an alternative scenario. In the IR these models are chirally broken with massless Goldstone pions. Coupling to the SM fields breaks the vacuum alignment and lifts some of the pion masses and the Higgs boson emerges as a pseudo Nambu-Goldstone boson (pNGB) \cite{Kaplan:1983fs,Kaplan:1983sm,Dugan:1984hq,Agashe:2004rs,Ferretti:2013kya,Ferretti:2014qta,Ma:2015gra,Vecchi:2015fma,Franzosi:2016aoo}.  Particularly well studied is the minimal scenario based on an SU(2) gauge theory with two Dirac flavors see e.g.~\cite{Hietanen:2014xca,Arthur:2016dir,DeGrand:2016mxr}.  Adding heavy  flavors will make both the dilaton-like and the pNGB Higgs systems conformal in the UV. The heavy mass  controls the``walking" behavior and  anomalous dimensions are determined by the IRFP. While the heavy flavors do not  effect  the  light-light spectrum strongly,   the  heavy-light and heavy-heavy resonances  would however be experimentally observable.

Fermion masses in both scenarios are generated either via 4-fermion interactions or partial compositeness. Both mechanisms might require UV properties similar to conformal systems as well. The new composite sector has to be coupled to the SM see e.g.~\cite{Ma:2015gra}. This leads to a radiative potential for the Higgs as pNGB and thus contributes to the mass of the Goldstone bosons. Other interactions like top-Yukawa couplings could also be significant.  In the following we focus on the new strongly interacting sector in isolation and leave investigations of couplings to the SM for future investigations.

\section{Hyperscaling in mass-split  systems}
\label{sec:hyperscaling}

 Hyperscaling in mass-split systems in the basin of attraction of a conformal IRFP follows from   Wilsonian renormalization group considerations.  For concreteness we assume lattice regularization and work with bare parameters that,  inside the conformal window, can be  separated into irrelevant gauge couplings $g_i$ and relevant lattice masses $\widehat m_i = a m_i$. The critical surface is given by $\widehat m_i = 0$ where the system is conformal at the IRFP  $g_i^\star$.

 In the vicinity of the IRFP an RG transformation that changes the scale $\mu \to \mu^\prime = \mu /b $ ($b>1$)  drives the gauge couplings to $g_i^\star$,  while masses transform with the scaling dimension $y_m=1+\gamma_m$ as $\widehat m_i \to \widehat m_i^\prime   = b^{y_{m}} \widehat m_i$ with $\gamma_m$ the universal anomalous dimension at the IRFP.
The correlation function of an operator $H$, after rescaling all dimensional quantities  by $b$,  change as
 \begin{align}
C_H(t; g_i, \widehat m_i,\mu) =  b^{-2y _H} C_H(t/b; g_i^\prime, \widehat m_i^\prime,\mu)\, ,
\label{eq:C_H1}
 \end{align} 
 where $y_H$ is the scaling dimension of $H$~\cite{DeGrand:2009mt,DelDebbio:2010ze}.
 As $b$ increases 
 the fermion mass  increases and  the fermions decouple from the IR dynamics around  $\widehat m_i^\prime = \mathcal{O}(1)$, i.e. when  the mass is above the cutoff. In RG language this is the scale identified as the IR scale, $\Lambda_{\rm{IR}}$.

  In our model we assume two different  fermion masses, $\widehat m_h = a m_h$ and $\widehat m_\ell = a m_\ell$, $\widehat m_h \ge \widehat m_\ell$.  Since both masses scale with the same exponent $y_m$, the dependence on $\widehat m_i^\prime = (\widehat m_h^\prime$, $\widehat m_\ell^\prime)$ in Eq.~(\ref{eq:C_H1}) can be replaced with $(\widehat m_h^\prime, \widehat m_\ell/ \widehat m_h) =( \widehat m_h^\prime, m_\ell/ m_h ) $
  \begin{align}
C_H(t; g_i, \widehat m_i,\mu) =  b^{-2y _H} C_H(t / b ; g_i^\prime, \widehat m_h^\prime  , m_\ell /  m_h, \mu).
\label{eq:C_H1b}
 \end{align} 
 
  The heavy fermions decouple when $\widehat m_h^\prime = b^{y_m} \widehat m_h = \mathcal{O}(1)$, and below that scale any dependence on $\widehat m_h$ is through the ratio $m_\ell/m_h$. We can identify this scale as the UV scale $\Lambda_{\rm{UV}}$ which is much lower than the cutoff scale set by the lattice spacing, $\Lambda_{\rm{cut}}\approx 1/a$.  The light flavors  still set the IR scale at $b = \widehat m_\ell^{-1/y_m}$ and  \eq{eq:C_H1b} reduces to
   \begin{align}
C_H(t; g_i, \widehat m_i,\mu) =  \widehat m_\ell^{2y _H/y_m} C_H(t \widehat m_\ell^{1/y_m} ; g_i^\prime ,  m_\ell /  m_h, \mu).
\label{eq:C_H1c} 
 \end{align} 
 
 Any  correlation function is expected to show exponential behavior at large distances,
  \begin{align}
 C_H(t; g_i,\widehat m_i,\mu) \propto e^{-M_H t}, \quad \quad t \to \infty.
 \label{eq:C_asymp} 
 \end{align}
 Comparing the $t$ dependence of  Eqs.~(\ref{eq:C_H1c}) and (\ref{eq:C_asymp}) leads to the  scaling relation
 \begin{align}
 a M_H = (\widehat m_\ell)^{1/y_m}  F_H( m_\ell/ m_h),
\label{eq:M_scaling_b}
 \end{align}
 where $ F_H$ is some  function of $ m_\ell/ m_h$ only, assuming $b$ is large enough that the gauge couplings take their IRFP value, $g^\prime_i = g^\star_i$. Ratios of masses
  \begin{align}
\frac{M_{H1} }{M_{H2}} = \frac{ F_{H1}( m_\ell/  m_h)}{  F_{H2}( m_\ell/ m_h)} 
\label{eq:R_scaling}
 \end{align}
depend only on $ m_\ell/  m_h$,  though the scaling function $F_{H1}/F_{H2}$ is different for different observables.
Between the IR and UV scales the system describes $N_\ell$ chirally broken fermions, yet the influence of the $N_h$ heavy flavors  is still evident though the universal dependence on $m_\ell/m_h$ and the scaling exponent $y_m$ in \eq{eq:M_scaling_b}. This scaling behavior is unlike in QCD and the consequence of the conformal IRFP that governs the system between the cutoff and UV scales. 

In our scaling tests we consider ratios of light-light and heavy-heavy hadrons and find that predictions  for different  $(m_\ell,m_h)$ fall on universal curves as  function of $ m_\ell/ m_h$. We expect the same to hold for the heavy-light spectrum but did not verify it by numerical simulations. Small deviations from universality can arise from corrections to scaling due to the slowly running gauge coupling, i.e.~deviations from $g_i^\prime = g_i^\star$. We have investigated these corrections within the $N_f=12$ system \cite{Cheng:2013xha}. 

Increasing $m_h$ will drive the system out of the basin of attraction of the IRFP. Scaling violations start to grow, higher order corrections contribute to $\gamma_m$ as well as Eqs.~(\ref{eq:M_scaling_b}) and (\ref{eq:R_scaling}). The gauge coupling becomes a relevant parameter, thus the functions $F_H$ depend on $g^2$, and the system becomes QCD-like.

\section{Numerical Results}
\label{sec:numres}

\begin{figure}[tb]
  \centering
 {\includegraphics[height=0.25\textheight]{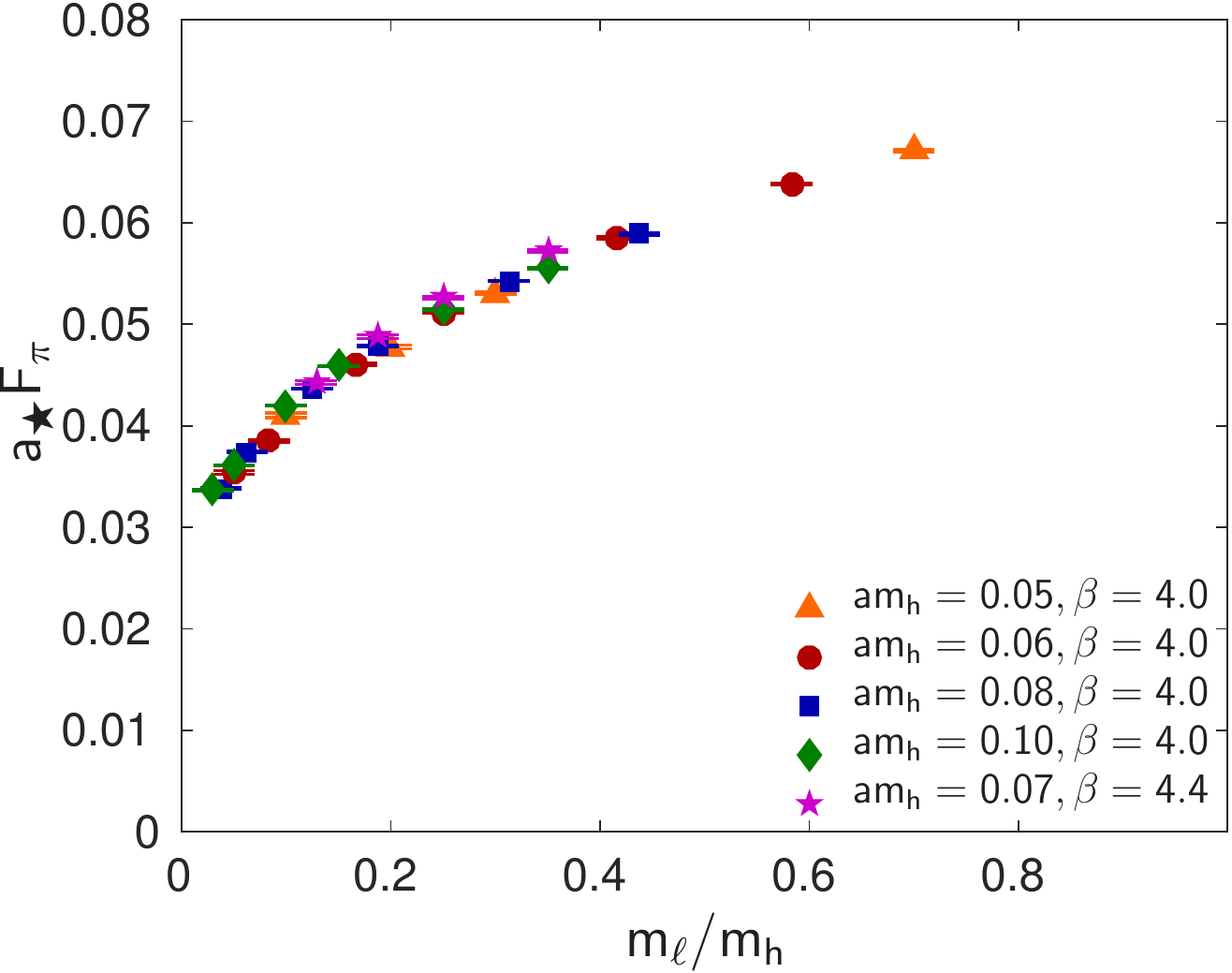}}  
  \caption{Dependence of  the light-light decay constant $F_\pi$ in units of $a_\bigstar$ on the hyperscaling variable $m_\ell / m_h$ (error bars are statistical only). Different colors and symbols correspond to different $\widehat m_h$ and $\beta$ values.}
  \label{fig:scale_fpi}
\end{figure}

\begin{figure*}[tb]
  \centering
  \parbox{0.16\columnwidth}{\includegraphics[height=0.25\textheight]{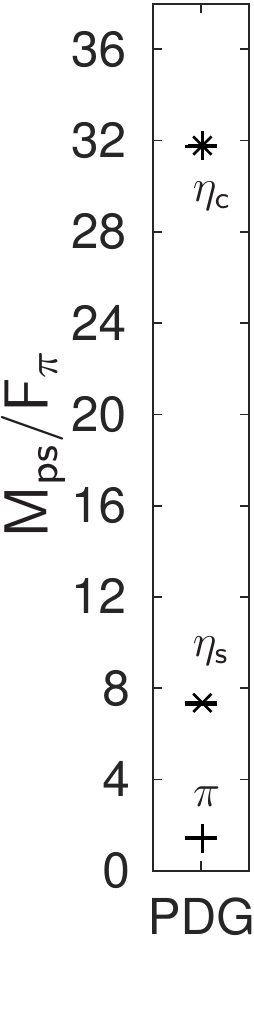}}  
  \parbox{0.38\columnwidth}{\includegraphics[height=0.25\textheight]{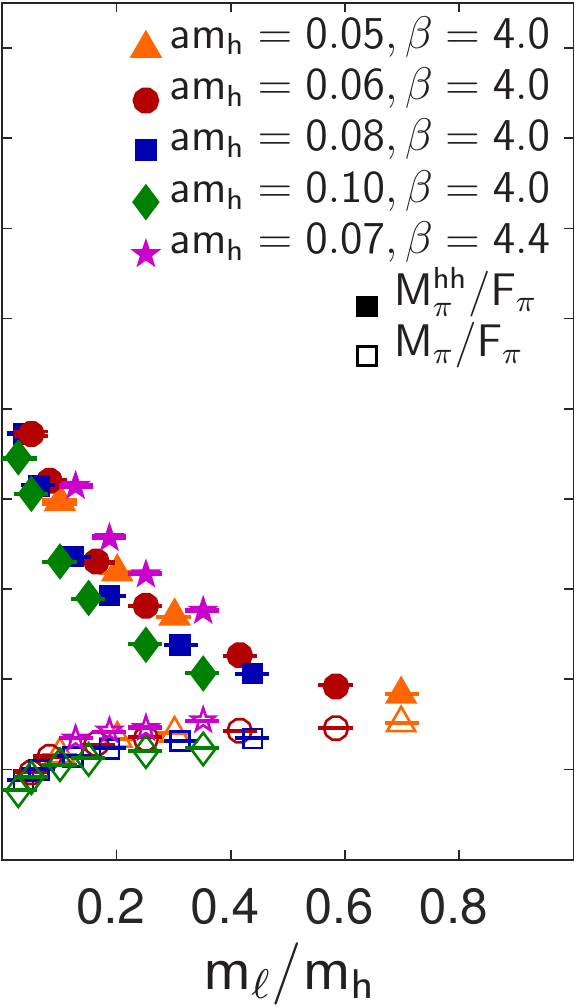}}
  \parbox{0.045\columnwidth}{\includegraphics[height=0.25\textheight]{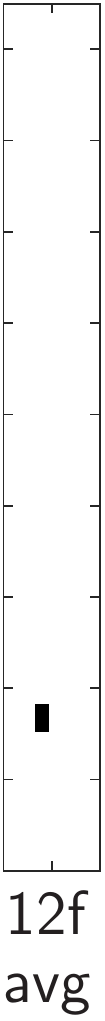}}
  \hspace{5mm}
  \parbox{0.16\columnwidth}{\includegraphics[height=0.25\textheight]{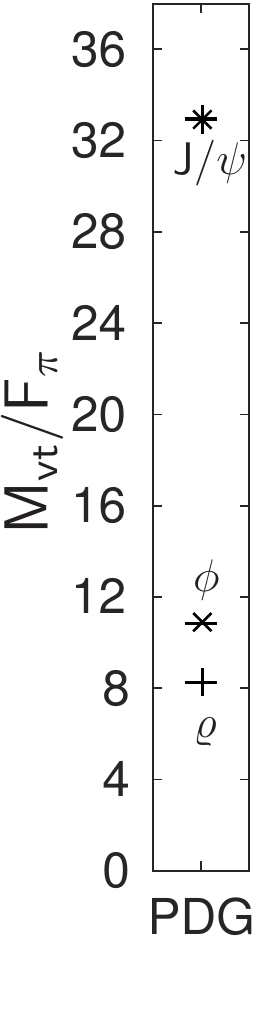}}  
  \parbox{0.38\columnwidth}{\includegraphics[height=0.25\textheight]{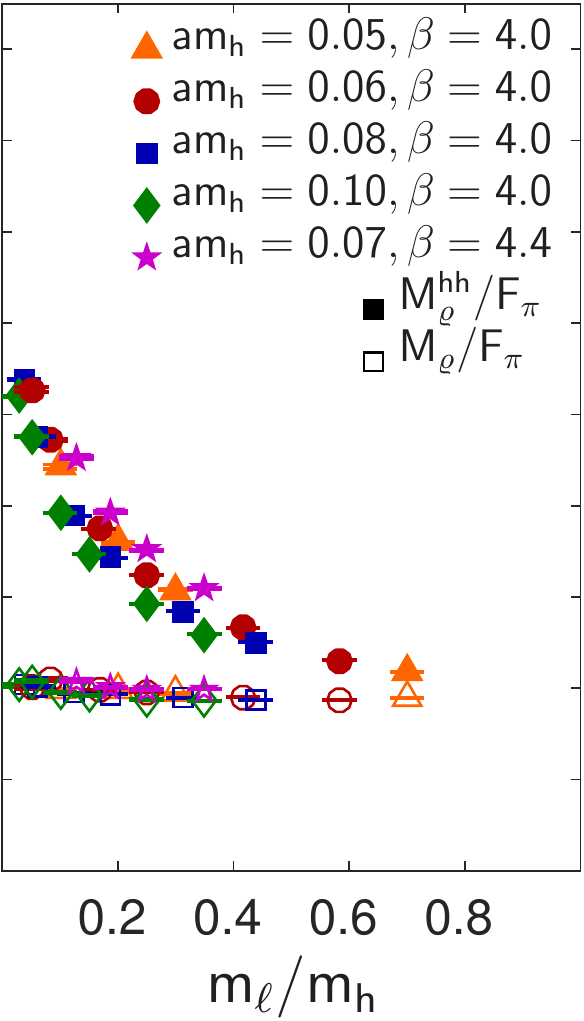}}
  \parbox{0.045\columnwidth}{\includegraphics[height=0.25\textheight]{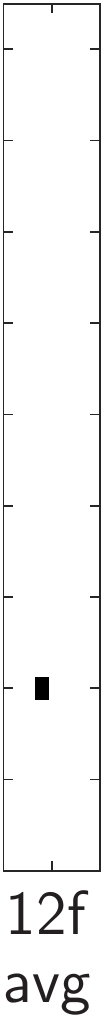}}
  \hspace{5mm}
  \parbox{0.16\columnwidth}{\includegraphics[height=0.25\textheight]{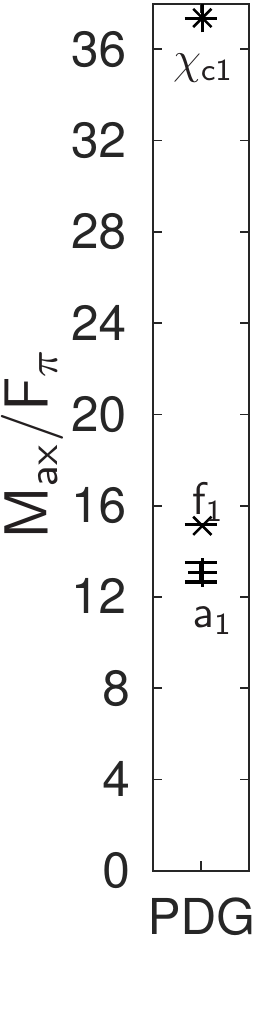}}  
  \parbox{0.38\columnwidth}{\includegraphics[height=0.25\textheight]{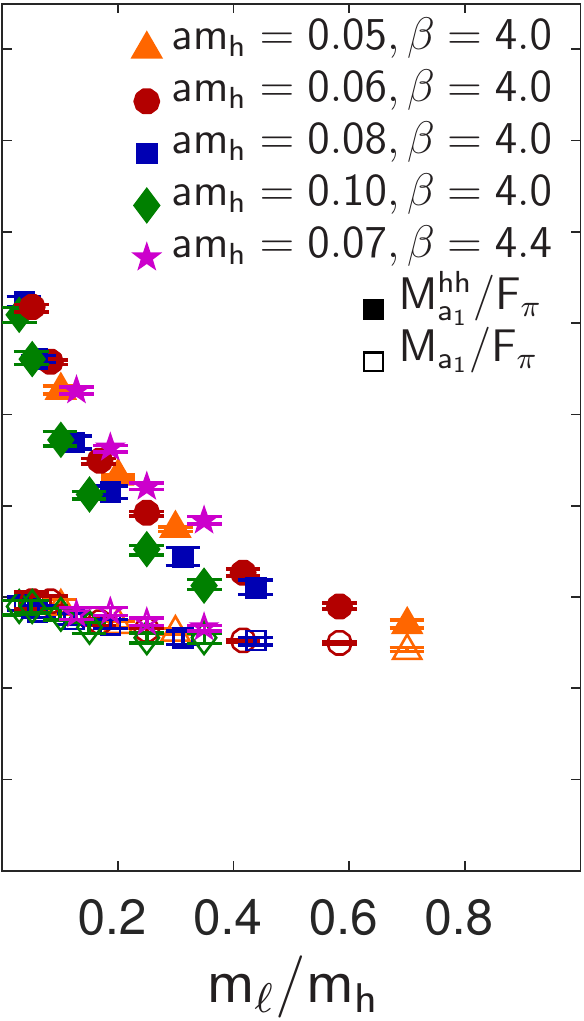}}
  \parbox{0.045\columnwidth}{\includegraphics[height=0.25\textheight]{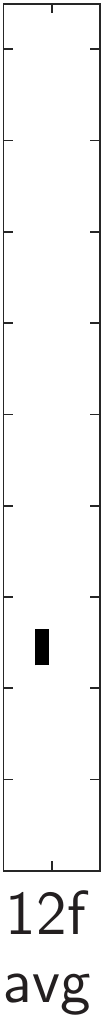}}  
  \caption{
    The three set of panels show dimensionless ratios for pseudoscalar (ps), vector (vt), and axial (ax) meson masses in units of $F_\pi$. The wide central panels show our data (with statistical errors only) as  function of $m_\ell/ m_h$. Different colors and symbols indicate the different $m_h$ and $\beta$ values, while filled (open) symbols denote states of the heavy-heavy (light-light) spectrum. The small panels to the right show averaged values for degenerate 12 flavors \cite{Aoki:2012eq,Fodor:2011tu,Cheng:2013xha,Aoki:2013zsa} and the panels on the left the corresponding PDG values \cite{Agashe:2014kda} for QCD divided by $F_\pi=94$ MeV. Values for the corresponding bottomonium states ($\eta_b$, $\Upsilon$, and $\chi_{b1}$) are too heavy to be shown on a reasonable scale. While the pseudoscalar and vector states are in general well understood in QCD, pure $(s\bar s)$ states do not occur in nature. For the $\eta_s$ mass, we use the lattice determination, $M_{\eta_s} = 688.5(2.2)$ MeV \cite{Dowdall:2013rya}, and quote for the vector and axial the PDG entries for the $\phi(1020)$ and $f_1(1420)$, respectively. Regardless of ambiguities in the QCD values, these plots highlight the different character of our heavy-heavy spectrum. Due to the presence of an IRFP, the system shows hyperscaling and we observe independence of the $m_h$, an unusual behavior in QCD standards.}
  \label{fig:ps_vt_ax}
\end{figure*}

Our lattice model is based on SU(3) gauge fields and four light and eight heavy flavors of fundamental fermions. We use  staggered fermions with nHYP smearing and the Wilson plaquette gauge action with fundamental and adjoint terms~\cite{Hasenfratz:2001hp,Hasenfratz:2007rf,Hasenfratz:2011xn}. In the  IR,  the system corresponds, e.g., to the composite Higgs model of Ref.~\cite{Ma:2015gra}, while in the UV it describes the conformal system of 12 degenerate flavors.\footnote{The $N_f=12$ flavor model has been investigated by different groups using lattice techniques \cite{Hasenfratz:2011xn,DeGrand:2011cu,Fodor:2011tu,Aoki:2012eq,Cheng:2013eu,Cheng:2013xha,Ishikawa:2013tua,Cheng:2014jba,Lombardo:2014pda,Lombardo:2015oha,Lin:2015zpa,Fodor:2016zil}. Most results are in agreement, concluding that $N_f=12$ is conformal but concerns were raised in Refs.~\cite{Fodor:2011tu,Fodor:2016zil}. A new step scaling study~\cite{Hasenfratz:2016dou} addresses those concerns and  identifies an infrared fixed point. Further, conformality is supported by a recent study of 10 fundamental flavors using domain-wall fermions that also identifies an IRFP \cite{Chiu:2016uui,Chiu:2017kza}. If $N_f=10$ is conformal, $N_f=12$ must be conformal, too.}

We investigate the model at two values of the gauge coupling $\beta$: At $\beta=4.0$ we ran simulations using four different  values of the heavy mass $\widehat m_h=0.05$, 0.06, 0.08 and 0.100,  while at $\beta=4.4$ we simulate with heavy mass $\widehat m_h=0.07$. For each heavy mass $\widehat m_h$, we simulate  4 to 6 $\widehat m_\ell$ values. Numerical simulations are carried out using FUEL/qhmc \cite{Osborn:2014kda,FUEL}.
In a conformal system the gauge coupling is irrelevant, the scale or lattice spacing depends on the fermion masses $\widehat m_\ell$ and $\widehat m_h$. We use the gradient flow scale \cite{Luscher:2010iy,Hasenfratz:2015xpa} to convert our data to the same lattice unit denoted by $a_\bigstar$ which we choose to match to the lattice spacing of the ensemble with $(\beta, \widehat m_\ell, \widehat m_h) = (4.0, 0.003, 0.080)$. For a first test on hyperscaling, we show $a_\bigstar F_\pi$ as function of the dimensionless ratio $m_\ell/m_h$ in Fig.~\ref{fig:scale_fpi}.  As predicted, the 26 independent ensembles --- corresponding to different $\widehat m_h$ and $\beta$ values ---  map out a unique trajectory.
The pion decay constant  is a particularly important quantity when considering the embedding of the Standard Model in a BSM system because $F_\pi$  is directly related to the $vev$ of the SM: $F_\pi = vev/\sin\chi$, where $\sin\chi$ is the vacuum alignment angle of composite Higgs systems ($\sin\chi=1$ in the dilaton-like scenario). The behavior of $F_\pi$ shown in \fig{fig:scale_fpi} not only supports hyperscaling, it also  demonstrates that the $4\ell+8h$ model is chirally broken with finite $F_\pi$ in the chiral limit.

We continue our study of hyperscaling by investigating  light-light and heavy-heavy pseudoscalar, vector, and axial resonances. Figure \ref{fig:ps_vt_ax} (wide panels) show dimensionless ratios of their masses in units of $F_\pi$  as function of $ m_\ell/ m_h$. As expected from hyperscaling, all states  at both gauge couplings follow unique curves. Considering the limit $m_\ell/ m_h \to 1$ (degenerate 12 flavors), both heavy-heavy and light-light values approach  the $N_f=12$ values \cite{Fodor:2011tu,Aoki:2012eq,Cheng:2013xha,Aoki:2013zsa} as depicted in the small panels to the right. Taking the limit $m_\ell/m_h \to 0$, we show for comparison PDG values \cite{Agashe:2014kda} for QCD on the small panels to the left including resonances dominated by $(s\bar s)$ and charmonium states. While the light-light states match the QCD values closely, the heavy-heavy spectrum is qualitatively different:  
\begin{itemize}
\item Due to the presence of an IRFP, ratios of heavy-heavy resonances exhibit hyperscaling and are independent of both the gauge coupling and  $m_h$\footnote{This is to be contrasted with  QCD  where heavy hadron masses are
approximately  proportional to the sum of the constituent quark masses, as
observed by experiment and also in lattice simulations.}
\item Heavy-heavy resonances exhibit a significant dependence on the light sea quark
mass 
\item The heavy-heavy resonances are only about a factor 2 -- 3 heavier than the light-light states 
\item Although not measured, heavy-light states are expected to show the same hyperscaling and lie between  the light-light and heavy-heavy spectrum
\end{itemize}

Small corrections to the overall behavior arise from scaling violations (corrections to scaling) due to lattice artifacts. Hyperscaling is only expected if  the irrelevant gauge couplings take their fixed point values, $g_i^\prime = g_i^\star$, in \eq{eq:C_H1c}. For slowly evolving gauge couplings this could require a large scale change since $b\approx \widehat m_h^{-1/y_m}$,  i.e.~small $m_h$ values. Corrections to scaling were investigated and found to be significant  for 12 degenerate flavors~\cite{Cheng:2013xha}. Further,  the 2-point functions leading to the data shown in Fig.~\ref{fig:ps_vt_ax} are subject to discretization errors of the fermion action. These discretization errors are known to grow for increasing quark masses. In case of the ratios over $F_\pi$, these errors largely cancel for the light-light resonances because discretization errors in numerator and denominator are similar, whereas in case of the heavy-heavy resonances this is not the case, leading to the somewhat larger  ``scatter" in the heavy-heavy data points.

\begin{figure}[tb]
  \centering
  \parbox{0.50\columnwidth}{\includegraphics[height=0.25\textheight]{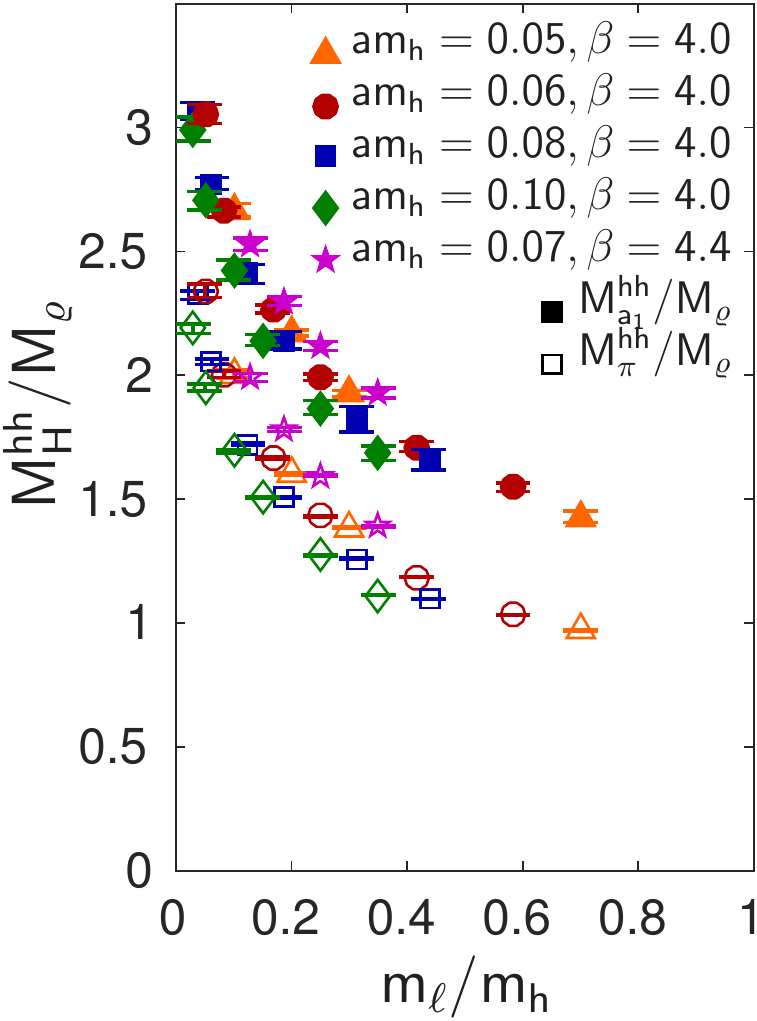}}  
  \parbox{0.48\columnwidth}{\includegraphics[height=0.25\textheight]{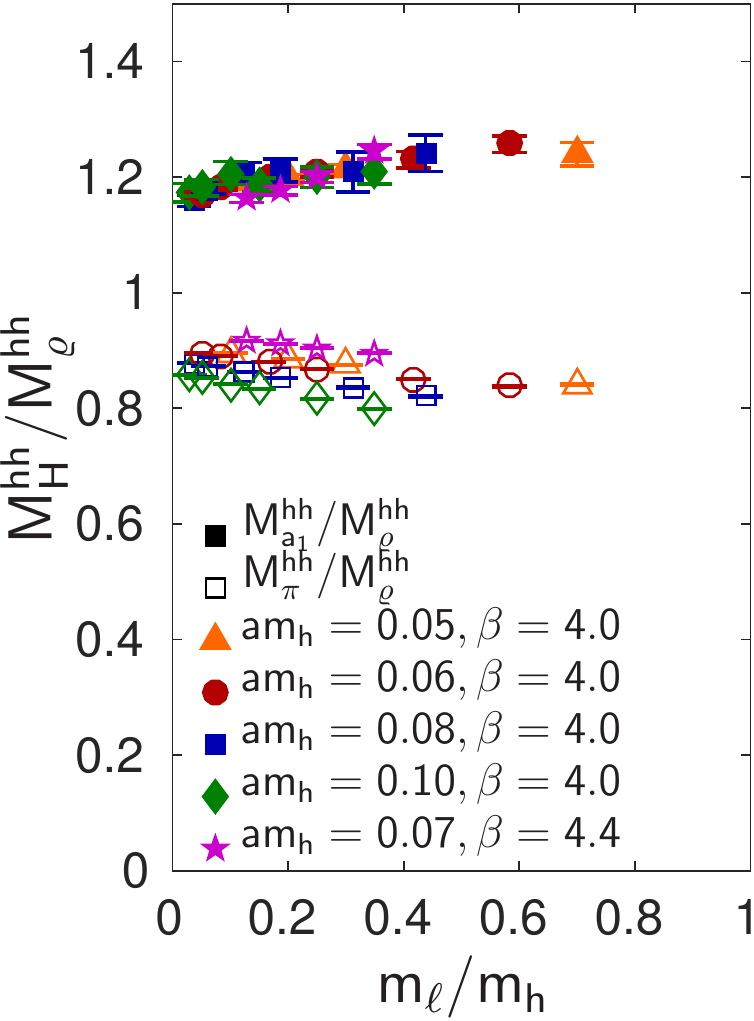}}
  \caption{Mass of the heavy-heavy pseudoscalar (pion) and axial (a1) shown as ratios over the light-light vector mass $M_\varrho$ (left) and the heavy-heavy vector mass $M_\varrho^{hh}$ (right) as function of $m_\ell/m_h$ (statistical errors only). Hyperscaling is present in both plots, but when dividing by $M_\varrho^{hh}$, the dependence on $m_\ell$ cancels almost entirely and discretization errors are reduced. }
  \label{fig:nucleon_ratios}
\end{figure}

To understand this better, we show in Fig.~\ref{fig:nucleon_ratios} the ratios of heavy-heavy pseudoscalar and axial masses  over the light-light rho mass (left plot) and the heavy-heavy rho mass (right plot).\footnote{Both $\varrho^{\ell\ell}$ and $\varrho^{hh}$ are stable since in our simulations they are energetically not allowed to decay to two pions.}  Both plots show hyperscaling in $m_h$. While the ratios over the light-light  $M_\varrho$ again introduces a strong dependence on $m_\ell$, the $m_\ell$ dependence is almost entirely canceled when plotting the ratios over $M_\varrho^{hh}$. In that case, also the scatter of the data points is reduced because now there is a better cancellation of discretization errors in the heavy-heavy quantities. However, some scaling violations, especially for the pseudoscalar state remain.

The curve collapse of the different $m_h$ spectra demonstrated in Figs. \ref{fig:ps_vt_ax} and \ref{fig:nucleon_ratios} is the consequence of hyperscaling at the conformal IRFP. We expect strong violation of this scaling as $m_h$ increases beyond the basin of attraction of the conformal  IRFP. Eventually the spectrum could become similar to QCD where  light and heavy flavor masses can be tuned independently around the perturbative Gaussian fixed point.  

\section{Outlook and Conclusion}

In this work we investigate systems where the fermions are split into $N_h$ heavy and $N_\ell$ light flavors. These models are examples for systems with large scale separation if the massless system with $N_f=N_\ell+N_h$ flavors is conformal, while  only $N_\ell$ massless flavors  are chirally broken. In the UV where all flavor masses are much lighter than the energy scale, the flavors can be considered degenerate and the dynamics is controlled by the conformal fixed point. Once the energy scale drops below the heavy flavor mass, the heavy flavors decouple and in the IR the system is chirally broken with only $N_\ell$ light flavors.  The scale separation is fully controlled by the  mass of the heavy flavors.

Hyperscaling relations at the conformal fixed point control the scaling behavior of  the light-light, heavy-light, and heavy-heavy resonance spectrum. General Wilsonian renormalization group considerations imply that dimensionless ratios depend only on the ratio of the flavor masses $m_\ell/m_h$ and not their individual values. This behavior is the consequence of the conformal fixed point and is very different from the well understood  QCD case. This property of mass-split systems is  general and  applies to  similar models.

We have verified the hyperscaling expectations in numerical simulations with our model of four light and eight heavy flavors. We found that the heavy-heavy spectrum is only a couple of times heavier than the light-light one and independent of the heavy fermion mass. This property of the heavy-heavy spectrum is fundamentally different  from what is observed in  QCD where quarkonia masses are proportional to the constituent quark mass. Furthermore, we observe that light-light but also heavy-heavy resonances are subject to large sea-quark mass effects; in QCD those effects are, in particular for heavy-heavy states, largely suppressed.

If this system describes BSM phenomenology with the light-light resonances in the few TeV range, the heavy-light and heavy-heavy states would be within the LHC range as well. Discriminating the various BSM models is however challenging because the light-light spectrum shows  little changes overall when varying the number of flavors or fermion representations \cite{ Aoki:2014oha,Aoki:2013xza, Appelquist:2016viq,Fodor:2014pqa,Brower:2015owo}. It is therefore interesting to investigate how the heavy-heavy or  heavy-light spectrum changes with the number of flavors that changes the anomalous dimension of the conformal fixed point in the UV.


\begin{acknowledgments}
The authors thank their colleagues in the LSD Collaboration for fruitful and inspiring discussions. 
Computations for this work were carried out in part on facilities of the USQCD Collaboration, which are funded by the Office of Science of the U.S.~Department of Energy, on computers at the MGHPCC, in part funded by the National Science Foundation (award OCI-1229059), and on computers allocated under the NSF Xsede program to the project TG-PHY120002. 
We thank Boston University, Fermilab, the NSF and the U.S.~DOE for providing the facilities essential for the completion of this work.  A.H. acknowledges support by DOE grant 
DE-SC0010005 and C.R. by DOE grant DE-SC0015845.   This project has received funding from the European Union's Horizon 2020 research and innovation programme under the Marie Sk{\l}odowska-Curie grant agreement No 659322. 
\end{acknowledgments}


\bibliography{../General/BSM}
\bibliographystyle{apsrev4-1} 

\end{document}